\documentclass[aps,prb,preprint,superscriptaddress]{revtex4-1}

\usepackage{graphicx}% Include figure files
\usepackage{dcolumn}% Align table columns on decimal point
\usepackage{bm}% bold math
\usepackage{multirow}

\begin{document}

\title{Comment on  ``Electron-phonon coupling in two-dimensional silicene and germanene" }

\author{M.E. Cifuentes-Quintal}
\email[]{cifuentes.quintal@gmail.com\\miguel.cifuentes@cinvestav.mx}
\affiliation{Departamento de F\'isica Aplicada, Centro de Investigaci\'on y de Estudios Avanzados del IPN,\\ Apartado Postal 73, Cordemex, 
97310 M\'erida, Yucat\'an, Mexico}

\author{O. de la Pe\~na-Seaman}
\affiliation{Instituto de F\'isica, Benem\'erita Universidad Aut\'onoma de Puebla,\\ Apartado postal J-48, 72570, Puebla, Puebla, Mexico}

\author{R. de Coss}
\affiliation{Departamento de F\'isica Aplicada, Centro de Investigaci\'on y de Estudios Avanzados del IPN,\\ Apartado Postal 73, Cordemex, 
97310 M\'erida, Yucat\'an, Mexico}

\begin{abstract}

In their work, Yan \textit{et al.} [Phys. Rev. B \textbf{88}, 121403 (2013)]  employing density functional 
perturbation theory (DFPT) calculations, demonstrate that silicene and germanene show weaker Kohn anomalies  
in the $\Gamma$-$E_g$ and $K$-$A_1$  phonon modes, compared to graphene. 
Furthermore, they compute the electron phonon (e-ph) coupling matrix elements using the  frozen phonon approach
and found that in silicene the average e-ph coupling matrix-element square over the Fermi surface, $\langle g_{{\bf q}\nu}^2\rangle_{F}$,
is about 50\% of those in graphene, but in germanene is weaker and nearly negligible.
However, Yan  \textit{et al.} argues that the smaller Fermi velocity in silicene compensates the
reduced $\langle g_{{\bf q}\nu}^2\rangle_{F}$, leading to phonon linewidths ($\gamma_{{\bf q}\nu}$) slightly larger than those in graphene.
In this Comment, we show that the DFPT and the frozen phonon results of Yan \textit{et al.}  for silicene are inconsistent.
Additionally, we have evaluated the e-ph coupling using direct DFPT calculations, analytical relations, and frozen phonon calculations, 
and we found systematically that $\langle g_{{\bf q}\nu}^2\rangle_{F}$ and $\gamma_{{\bf q}\nu}$ 
in silicene are one order of magnitude smaller than in graphene, in contrast to the conclusions of Yan \textit{et al.}

\end{abstract}

\pacs{63.20.kd, 61.46.-w, 78.30.-j, 71.15.Mb} 
 
\maketitle

\date{\today}

In reference \onlinecite{yan2013}, Yan  \textit{et al.} report a first-principles study of the Kohn anomalies
and electron-phonon (e-ph) coupling in low buckled monolayer silicene and germanene. 
The phonon frequencies and associated eigenvectors were computed using the density functional perturbation 
theory (DFPT), and the e-ph coupling matrix elements by a frozen-phonon approach (FPA),
which was previously developed by the same authors for graphene,\cite{yan2009}
with results in very good agreement with other calculations.\cite{piscanec2004,lazzeri2006,cifuentes2016}
The phonon frequency shift and the linewidth were obtained  from the phonon self-energy
within the Migdal approximation.

By varying the electronic smearing occupation in the phonon calculations for silicene and germanene, 
Yan \textit{et al.}  demonstrate the presence of Kohn anomalies in the highest optical branch 
of the ${\bf q}\nu$ phonon modes, i.e. $\Gamma$-$E_g$ and $K$-$A_1$.
However, the range of frequency variation as a function of the smearing is significantly smaller than 
 in graphene, implying a much weaker e-ph coupling in silicene and germanene.
In order to perform a quantitative comparison, Yan \textit{et al.} compute the
average e-ph coupling matrix-element square over the Fermi surface ($\langle g_{{\bf q}\nu}^2\rangle_{F}$)
for the $\Gamma$-$E_g$ and $K$-$A_1$ phonon modes.
In silicene, they found that $\langle g_{{\bf q}\nu}^2\rangle_{F}$ is about 50\% of those in graphene,
while in germanene is weaker and nearly negligible.
Finally, Yan  \textit{et al.} argue that the smaller Fermi velocity in silicene compensates the
reduced $\langle g_{{\bf q}\nu}^2\rangle_{F}$, leading to phonon linewidths ($\gamma_{{\bf q}\nu}$)  
slightly larger than those in graphene.

In this Comment, we show that for silicene, the FPA results of $\langle g_{{\bf q}\nu}^2\rangle_{F}$  reported by Yan \textit{et al.}
are inconsistent with their DFPT phonon dispersion, which gives rise to an artificially enhanced $\gamma_{{\bf q}\nu}$.
In addition, by computing $\langle g_{{\bf q}\nu}^2\rangle_{F}$ employing direct DFPT calculations, analytical relations, and frozen phonon calculations,
we found systematically that the e-ph coupling in silicene is one order of magnitude smaller than in graphene,
in contrast to the original conclusions of Yan \textit{et al.}

We begin our analysis by briefly recovering the seminal work of Piscanec \textit{et al.},\cite{piscanec2004} 
which demonstrates that in graphene is possible to obtain  the e-ph coupling
entirely from the knowledge of the electronic band structure and phonon dispersion.
In that work, it was shown that
the slope $\alpha_{{\bf q}\nu}$ of the phonon branches  around the Kohn anomalies  in the $\Gamma$-$E_{2g}$ and $K$-$A'_1$ 
phonon modes  is proportional to the ratio between $\langle g_{{\bf q}\nu}^2\rangle_{F}$ and 
the slope $\beta$ of the electronic bands near the Fermi level:
%%%%%%%%%%%%%%%%%%%%%%%%%%%%%%%%%%%%%%%%%%%%%%%%%%%%%%%%%%%%%%%%%%%%%%%%%%%%%%%%%%%%%%%%%%%%%%%
\begin{equation}
\label{alpha}
\alpha_{{\bf q}\nu} = \frac{\sqrt3 \pi^2}{\beta}  \langle g_{{\bf q}\nu}^2\rangle_{F} ,
\end{equation}
%%%%%%%%%%%%%%%%%%%%%%%%%%%%%%%%%%%%%%%%%%%%%%%%%%%%%%%%%%%%%%%%%%%%%%%%%%%%%%%%%%%%%%%%%%%%%%%
where $\beta$ and $\alpha_{{\bf q}\nu}$ are given in energy units,  due to the momentum space was expressed in units of  $2\pi/a$,
being $a$ the lattice constant. 
In a subsequent work,\cite{lazzeri2006} it was also demonstrated that the phonon linewidth of the $\Gamma$-$E_{2g}$ phonon mode is:
%%%%%%%%%%%%%%%%%%%%%%%%%%%%%%%%%%%%%%%%%%%%%%%%%%%%%%%%%%%%%%%%%%%%%%%%%%%%%%%%%%%%%%%%%%%%%%%
\begin{equation}
\label{gamma}
\gamma_{\Gamma}^{ } = \frac{2 \sqrt{3} \pi^2 \omega_{\Gamma}^{ }}{\beta^2 } \langle g_{{\bf q}\nu}^2\rangle_{F} = \frac{ 2 \alpha_{\Gamma}^{ } \omega_{\Gamma}^{ }}{\beta },
\end{equation}
%%%%%%%%%%%%%%%%%%%%%%%%%%%%%%%%%%%%%%%%%%%%%%%%%%%%%%%%%%%%%%%%%%%%%%%%%%%%%%%%%%%%%%%%%%%%%%%
where $\omega_{\Gamma}^{}$ is the phonon frequency.
Thus, by extracting $\alpha_{{\bf q}\nu}$ and $\beta$ from their calculations and experimental results,\cite{maultzsch2004} 
Piscanec \textit{et al.} demonstrate that the results obtained with Eqs. \ref{alpha} and \ref{gamma}, 
are in good agreement with direct  DFPT calculations of $\langle g_{{\bf q}\nu}^2\rangle_{F}$.

It is important to note that in the derivation of Eqs. \ref{alpha} and \ref{gamma}, it was considered a conic shape for the electronic bands near the Fermi level,
a common feature in graphene, silicene, and germanene.
\footnote{At least in absence of spin-orbit coupling, as was considered in the article of Yan \textit{et al.}\cite{yan2013}}
Evenmore, the projection of the $\Gamma$-$E_{g}$ and $K$-$A_1$ phonon modes for the buckled structure in the hexagonal plane corresponds to the 
$\Gamma$-$E_{2g}$ and $K$-$A'_1$ phonon modes, respectively.  Thus, it could be expected that Eqs. \ref{alpha} and \ref{gamma} hold for silicene and germanene.

\begin{figure}
\includegraphics*[scale=1.0]{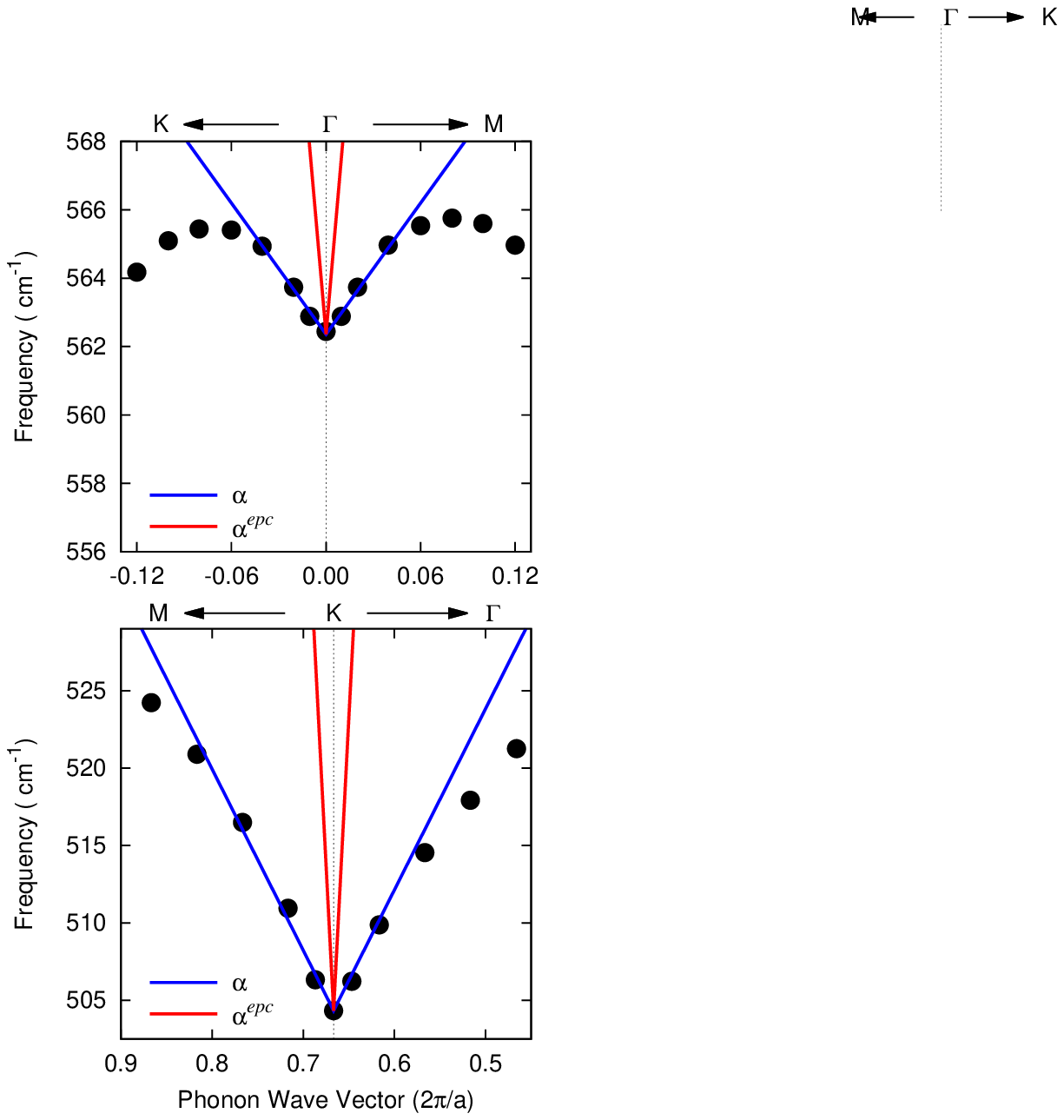}
\caption{ 
Kohn anomalies for silicene around the $\Gamma$ point (top) and the $K$ point (bottom).
The solid symbols are phonon frequencies obtained from Fig. 2 in Ref. [\onlinecite{yan2013}], for a smearing parameter of 0.005 Ry.
The blue line represents a linear fit to the phonon frequencies. 
The red line is the slope defined in Eq. \ref{alpha} and obtained using the values of $\beta$ and $\langle g_{{\bf q}\nu}^2\rangle_{F}$ 
reported by Yan \textit{et al.}\cite{yan2013}
}
\label{freq}
\end{figure}

In order to evaluate the agreement between the phonon dispersion and the e-ph coupling calculations of  Yan \textit{et al.}, 
we use their reported values for the Fermi velocity $v_F$ ($\beta=h v_F/a$) and $\omega_{\Gamma}^{}$, and then
we perform two consistency tests.
In the first one, by applying a linear fitting to the phonon frequencies of Fig. 2 in Ref. [\onlinecite{yan2013}], 
we obtain the slope $\alpha$ of the phonon branches around the Kohn anomalies (see Fig. \ref{freq}).
Then, we evaluate the Eqs. \ref{alpha} and \ref{gamma} to get $\langle g_{{\bf q}\nu}^2\rangle_{F}^{ph}$ and $\gamma_\Gamma^{ph}$,
which represent estimated values from the phonon dispersion.
In the second consistency test, using the reported values of Yan \textit{et al.}  for $\langle g_{{\bf q}\nu}^2\rangle_{F}$, 
we obtain $\alpha^{epc}$ and $\gamma_\Gamma^{epc}$, which correspond to estimated values from the e-ph coupling data.

%%%%%%%%%%%%%%%%%%%%%%%%%%%%%%%%%%%%%%%%%%
%%%%%%%%%%%%%%%%%%%%%%%%%%%%%%%%%%%%%%%%%%%%%%

In Table \ref{yan-values}, we summarize the phonon and e-ph coupling results of Yan \textit{et al.}, 
where for comparison we include the results of the two consistency tests  described above.
For silicene, the difference between the reported values for $\langle g_{{\bf q}\nu}^2\rangle_{F}$ and $\gamma_\Gamma$,
and the estimated ones $\langle g_{{\bf q}\nu}^2\rangle_{F}^{ph}$ and $\gamma_\Gamma^{ph}$  is remarkable. 
Furthermore, as shown in Fig. \ref{freq}, the slope $\alpha^{epc}$ is completely out of scale with respect to the
phonon dispersion reported by Yan \textit{et al.}
In fact, it is important to note that the values for $\alpha^{epc}$ in silicene at $\Gamma$ ($527$ cm$^{-1}$) and $K$ ($1130$ cm$^{-1}$)
are larger than in graphene\cite{piscanec2004} ($\alpha_\Gamma=396$ cm$^{-1}$ and $\alpha_K=973$ cm$^{-1}$),
indicating that the Kohn anomalies in silicene are stronger than in graphene.
However, the values $\gamma_\Gamma$ and $\gamma_\Gamma^{epc}$ show a very good agreement.
On the other hand, for germanene there is a better consistency among all the analyzed parameters.
For example,  the phonon frequencies obtained using the slopes $\alpha$ and $\alpha^{epc}$ differs  only $0.48$ cm$^{-1}$
at ${\bf q}= 0.03$ in the direction $\Gamma \rightarrow K$, close to the Kohn anomaly at $\Gamma$-$E_g$.
Contrary to our analysis, in their article Yan \textit{et al.} indicates that the computed $\langle g_{{\bf q}\nu}^2\rangle_{F}$
in silicene is consistent with the direct phonon calculations, and that surprisingly in germanene the e-ph coupling is negligible.

%%%%%%%%%%%%%%%%%%%%%%%%%%%%%%%%%%%%%%%%%%%%%%%%%%%%%%%%%%%%%%%%%%%%%%%%%%%%%%%%%%%%%%%%%%%%%%%%%%%%%%%%%%%%%%%%%%%%%%%%%%%%%%%%
\begin{table*}[!ht]
\centering
\caption{Summary of the results from Yan \textit{et al.},\cite{yan2013} and the two consistency tests.
The superindex stands for parameters calculated with Eq. \ref{alpha} and \ref{gamma}, ($ph$) using a linear fit to the  
the phonon dispersion, and ($epc$) using the reported values of $\langle g_{{\bf q}\nu}^2\rangle_{F}$.}
\begin{ruledtabular}
\begin{tabular}{lcccrcrccccccccc}

& 
\multicolumn{1}{c}{$\beta$} & 
${\bf q}$ & 
\multicolumn{1}{c}{$\omega$}& 
\multicolumn{1}{c}{$\alpha$}&
\multicolumn{1}{c}{$\langle g^2 \rangle_{F}$}&
\multicolumn{1}{c}{$\gamma$}& 
\multicolumn{1}{c}{$\langle g^2 \rangle_{F}^{ph}$}&
\multicolumn{1}{c}{$\gamma^{ph}$}& 
\multicolumn{1}{c}{$\alpha^{epc}$}&
\multicolumn{1}{c}{$\gamma^{epc}$}&
\\
& 
\multicolumn{1}{c}{(eV)}&
&
\multicolumn{1}{c}{(cm$^{-1}$)}&
\multicolumn{1}{c}{(cm$^{-1}$)}&
\multicolumn{1}{c}{(eV$^2$)}&
\multicolumn{1}{c}{(cm$^{-1}$)}&
\multicolumn{1}{c}{(eV$^2$)}&
\multicolumn{1}{c}{(cm$^{-1}$)}&
\multicolumn{1}{c}{(cm$^{-1}$)}&
\multicolumn{1}{c}{(cm$^{-1}$)}&
\\
\hline
\multirow{2}{*}{Silicene}&\multirow{2}{*}{5.83}& $\Gamma$ &562 & 64   & 0.0223 & 13.3 & 0.0027 & 1.5& 527  &12.6\\
                         &                     & $K$      &506 & 117  & 0.0478 & 21.5 & 0.0049 & -  & 1130 & - \\      
\hline
\multirow{2}{*}{Germanene}&\multirow{2}{*}{5.54}&$\Gamma$ &303 & 37   & 0.0021 & 0.6 & 0.0015 & 0.5& 52 & 0.7  \\
                          &                     &$K$      &267 & 73   & 0.0046 & 1.2 & 0.0030 &  - &114 & - \\
\end{tabular}
\end{ruledtabular}
\label{yan-values}
\end{table*}
%%%%%%%%%%%%%%%%%%%%%%%%%%%%%%%%%%%%%%%%%%%%%%%%%%%%%%%%%%%%%%%%%%%%%%%%%%%%%%%%%%%%%%%%%%%%%%%%%%%%%%%%%%%%%%%%%%%%%%%%%%%%%%%%

Although  numerical differences between the evaluation of Eqs. \ref{alpha} and \ref{gamma} with
respect to direct calculations of the phonon dispersion or the e-ph coupling are expected, our analysis indicates two possibilities:
the analytic relations developed for graphene does not hold for silicene, or there is a mistake in either the phonon 
or e-ph coupling calculations of Yan \textit{et al.}
Thus, we decide to perform  DFT-based first principles calculations of the phonon dispersion and e-ph coupling
in silicene and germanene, and then compare with the results of Yan \textit{et al.}

In order to clarify the inconsistent results of Yan \textit{et al.},\cite{yan2013} 
we try to keep our calculations as close as possible to that work.
Thus, we employ the plane-waves and pseudopotential (PWPP) method, as implemented in the Quantum ESPRESSO code,\cite{QE-2009}
and the local density approximation for the  exchange-correlation functional.\cite{PZ}
Core electrons were replaced by ultra-soft pseudopotentials taken from the PSLibrary,\cite{PSLIB} 
and the valence wave functions (charge density) were expanded in plane waves with a kinetic energy cut-off of 30 (360) Ry for silicene,
and 60 (600) Ry for germanene. 
Dynamical matrices and e-ph coupling matrix elements were computed by means of DFPT.
However, Yan \textit{et al.} failed to report the $k-$grid employed in their phonon calculations, a critical numerical parameter.
For that reason, after convergence tests, we employ a $72\times72$ $k-$grid for a Methfessel-Paxton\cite{MP} smearing of 0.005 Ry.
The e-ph coupling matrix elements were computed in a serie of $k-$grids from $576\times576$ up to $1008\times1008$, and the
electronic smearing varying from 0.0002 to 0.005 Ry. Finally, phonon linewidths were calculated using Eq. 3 of Yan \textit{et al.}
using a temperature of 15 K in the Fermi-Dirac occupation. 
Employing DFPT calculations, 
we have previously\cite{cifuentes2016} obtained $\langle g_{{\bf q}\nu}^2\rangle_{F}$ for graphene,
and that result is in very good agreement with Piscanec \textit{et al.}\cite{piscanec2004}
and the previous work of Yan \textit{et al.} for graphene in Ref. [\onlinecite{yan2009}].

In Table \ref{our-values}, we report our DFPT results of the phonon dispersion and e-ph coupling for silicene and germanene, 
 as well the two consistency tests previously applied to the data of Yan \textit{et al.}
For easy comparison and analysis, we use the same format as in Table I. %, which contains the results of Yan \textit{et al.}
We found that our phonon results ($\omega$ and $\alpha$) are in very good agreement with the work of Yan \textit{et al.},
but our direct DFPT calculations of $\langle g^2 \rangle_{F}$ and $\gamma$ are one order of magnitude lower than their reported values.
However, it is important to note that the evaluation of Eqs. \ref{alpha} and \ref{gamma} within our DFPT calculations give consistent results
for both silicene and germanene, indicating that something is missed in the results of  Yan \textit{et al.} for silicene.
Evenmore, the results for $\langle g_{{\bf q}\nu}^2\rangle_{F}^{ph}$ and $\gamma_\Gamma^{ph}$ obtained from the estimated value of $\alpha$ from 
the phonon dispersion of  Yan \textit{et al.} are in very good agreement with our corresponding results, clearly indicating that the 
frozen phonon calculation of Yan \textit{et al.} for the e-ph coupling matrix elements in silicene  is wrong.

%%%%%%%%%%%%%%%%%%%%%%%%%%%%%%%%%%%%%%%%%%%%%%%%%%%%%%%%%%%%%%%%%%%%%%%%%%%%%%%%%%%%%%%%%%%%%%%%%%%%%%%%%%%%%%%%%%%%%%%%%%%%%%%%
\begin{table*}
\centering
\caption{
Summary of our  DFPT results for the phonon dispersion and e-ph coupling in silicene and germanene, and the two consistency tests.
The superindex stands for parameters calculated with Eq. \ref{alpha} and \ref{gamma}, 
($ph$)  using the slope $\alpha$, and ($epc$) using the obtained value of $\langle g_{{\bf q}\nu}^2\rangle_{F}$.}
\begin{ruledtabular}
\begin{tabular}{lcccrcrccccccccc}
& 
\multicolumn{1}{c}{$\beta$} & 
${\bf q}$& 
\multicolumn{1}{c}{$\omega$}& 
\multicolumn{1}{c}{$\alpha$}&
\multicolumn{1}{c}{$\langle g^2 \rangle_{F}$}&
\multicolumn{1}{c}{$\gamma$}& 
\multicolumn{1}{c}{$\langle g^2 \rangle_{F}^{ph}$}&
\multicolumn{1}{c}{$\gamma^{ph}$}& 
\multicolumn{1}{c}{$\alpha^{epc}$}&
\multicolumn{1}{c}{$\gamma^{epc}$}&
\\
& 
\multicolumn{1}{c}{(eV)}&
&
\multicolumn{1}{c}{(cm$^{-1}$)}&
\multicolumn{1}{c}{(cm$^{-1}$)}&
\multicolumn{1}{c}{(eV$^2$)}&
\multicolumn{1}{c}{(cm$^{-1}$)}&
\multicolumn{1}{c}{(eV$^2$)}&
\multicolumn{1}{c}{(cm$^{-1}$)}&
\multicolumn{1}{c}{(cm$^{-1}$)}&
\multicolumn{1}{c}{(cm$^{-1}$)}&
\\
\hline
\multirow{2}{*}{Silicene}&\multirow{2}{*}{5.84}& $\Gamma$ &561 & 64   & 0.0031 &  1.8 & 0.0027 & 1.8& 78 & 1.0\\
                         &                     & $K$      &504 & 117  & 0.0063 &  3.3 & 0.0049 & -  & 148 & -   \\
\hline
\multirow{2}{*}{Germanene}&\multirow{2}{*}{5.61}&$\Gamma$ &309 & 35   & 0.0017 & 0.6 & 0.0015 & 0.5 & 41 & 0.6 \\
                          &                     &$K$      &278 & 78   & 0.0037 & 1.1 & 0.0030 &  -  & 91 & -   \\
\end{tabular}
\end{ruledtabular}
\label{our-values}
\end{table*}
%%%%%%%%%%%%%%%%%%%%%%%%%%%%%%%%%%%%%%%%%%%%%%%%%%%%%%%%%%%%%%%%%%%%%%%%%%%%%%%%%%%%%%%%%%%%%%%%%%%%%%%%%%%%%%%%%%%%%%%%%%%%%%%%

To rule out that the difference between our results and those of Yan \textit{et al.}  comes from numerical sources
such as the use of different pseudopotentials or $k-$grids, we perform frozen phonon calculations employing the
full-potential linearized augmented plane-wave with local orbitals (FP-LAPW+lo) method, as implemented in the ELK code.\cite{ELK} 
We use a muffin-tin radii ($R_{MT}$) of 2.11 and 2.21 $a.u.$ for silicene and germanene, respectively, and a $R_{MT}\times|G+k|_{MAX}=8.0$.
The angular momentum cut-off for the muffin-tin charge density and potential were expanded in crystal harmonics up to $l=8$.

For the FPA calculations, the atomic positions were displaced according to the $\Gamma$-$E_g$ and $K$-$A_1$ phonon modes by a small
distance $d$ of up to $0.010$ \AA. 
The phonon frequencies were computed by a quadratic fitting of the electronic total energy as a function of the atomic displacement, 
as in our previous works.\cite{omar2007,edgar}
Then, from the electronic band strucure we obtain the band gap $\Delta E$ for each displacement, and 
for the evaluation of  $\langle g_{{\bf q}\nu}^2\rangle_{F}$ we use the  the following relations previously 
reported by Lazzeri \textit{et al.}\cite{lazzeri2008} for graphene:
%%%%%%%%%%%%%%%%%%%%%%%%%%%%%%%%%%%%%%%%%%%%%%%%%%%%%%%%%%%%%%%%%%%%%%%%%%%%%%%%%%%%%%%%%%%%%%%
\begin{eqnarray}
\label{fp-g}
\langle g_{\Gamma}^2\rangle_{F} = \lim_{d\rightarrow 0} \frac{\hbar}{32 M \omega_{{\Gamma}} } \left(\frac{\Delta E}{d} \right)^2 \nonumber\\
\langle g_{K}^2\rangle_{F} = \lim_{d\rightarrow 0} \frac{\hbar}{16 M \omega_{K}} \left(\frac{\Delta E}{d} \right)^2,
\end{eqnarray}
%%%%%%%%%%%%%%%%%%%%%%%%%%%%%%%%%%%%%%%%%%%%%%%%%%%%%%%%%%%%%%%%%%%%%%%%%%%%%%%%%%%%%%%%%%%%%%%
where $M$ is the atomic mass. In the case of the $K$-$A_1$ phonon mode we use a $\sqrt{3}\times\sqrt{3}$ supercell, where the $K$ point and the
band gap is refolded in the $\Gamma$ point.

The results of our frozen phonon calculations are presented in Table \ref{table-fp}, 
where for comparison we have included the FPA results obtained using the PWPP method.
It is clear that the effects of using
different DFT methods for the calculation of the studied parameters  are practically negligibles. 
It is important to note that our frozen phonon results for $\langle g^2 \rangle_{F}$ are
in good agreement with the values in Table \ref{our-values} obtained from DFPT calculations and
the analytical relation of Eq. \ref{alpha}, indicating consistency and giving support to our results.

%%%%%%%%%%%%%%%%%%%%%%%%%%%%%%%%%%%%%%%%%%%%%%%%%%%%%%%%%%%%%%%%%%%%%%%%%%%%%%%%%%%%%%%%%%%%%%%%%%%%%%%%%%%%%%%%%%%%%%%%%%%%%%%%
\begin{table}
\centering
\caption{
Phonon frequencies, $\Delta E/d$, and $\langle g^2 \rangle_{F}$ for silicene, calculated using the  frozen phonon approximation, by means of two different DFT methods:
plane-waves and pseudopotential (PWPP), and the full-potential linearized augmented-plane wave with local orbitals (FP-LAPW+lo).}
\begin{ruledtabular}
\begin{tabular}{cccccc}
\multirow{2}{*}{Method} & 
\multirow{2}{*}{${\bf q}$} & 
\multicolumn{1}{c}{$\omega$}& 
\multicolumn{1}{c}{$\Delta E/d$}&
\multicolumn{1}{c}{$\langle g^2 \rangle_{F}$}&
\\
&
&
\multicolumn{1}{c}{(cm$^{-1}$)}&
\multicolumn{1}{c}{(eV/\AA)}&
\multicolumn{1}{c}{(eV$^2$)}&
\\
\hline
\multirow{2}{*}{PWPP}      & $\Gamma$ & 561 & 6.83  & 0.0031 \\
                           & $K$      & 504 & 6.54  & 0.0064 \\      
\multirow{2}{*}{FPLAPW+lo} & $\Gamma$ & 566 & 6.89  & 0.0031 \\
                           & $K$      & 510 & 6.62  & 0.0064 \\
\end{tabular}
\end{ruledtabular}
\label{table-fp}
\end{table}
%%%%%%%%%%%%%%%%%%%%%%%%%%%%%%%%%%%%%%%%%%%%%%%%%%%%%%%%%%%%%%%%%%%%%%%%%%%%%%%%%%%%%%%%%%%%%%%%%%%%%%%%%%%%%%%%%%%%%%%%%%%%%%%%

Finally, we want to comment that is well established that FPA and DFPT calculations should agree when those are performed carefully 
under the harmonic approximation, as we found  between our DFPT and FPA results. 
Thus, in this work we are not criticizing the use of the FPA to compute the e-ph coupling in silicene. 
However, given the above explained consistence between the three methodologies to compute the e-ph coupling, 
even when using two DFT codes with different approaches, 
it is clear that the FPA implementation of Yan \textit{et al.} has some mistake in the case of silicene.

In conclusion, in this comment we show that the result of Yan \textit{et al.}\cite{yan2013} for e-ph coupling in silicene is wrong.
From direct DFPT calculations, evaluation of analytical relations, and frozen phonon calculations, we found  systematically that the e-ph coupling in 
silicene is one order of magnitude smaller than in graphene.

\begin{acknowledgments}
The authors thank to Rolf Heid for a critical reading of the manuscript.
One of the authors (M.E.C.-Q.) gratefully acknowledges a student grant from CONACYT-M\'exico.
Computational resources was provided by ``Cluster H\'ibrido de Superc\'omputo, Xiuhcoatl" at
Cinvestav.
\end{acknowledgments}

%\bibliography{cifuentes_comment}

\begin{thebibliography}{15}%
\makeatletter
\providecommand \@ifxundefined [1]{%
 \@ifx{#1\undefined}
}%
\providecommand \@ifnum [1]{%
 \ifnum #1\expandafter \@firstoftwo
 \else \expandafter \@secondoftwo
 \fi
}%
\providecommand \@ifx [1]{%
 \ifx #1\expandafter \@firstoftwo
 \else \expandafter \@secondoftwo
 \fi
}%
\providecommand \natexlab [1]{#1}%
\providecommand \enquote  [1]{``#1''}%
\providecommand \bibnamefont  [1]{#1}%
\providecommand \bibfnamefont [1]{#1}%
\providecommand \citenamefont [1]{#1}%
\providecommand \href@noop [0]{\@secondoftwo}%
\providecommand \href [0]{\begingroup \@sanitize@url \@href}%
\providecommand \@href[1]{\@@startlink{#1}\@@href}%
\providecommand \@@href[1]{\endgroup#1\@@endlink}%
\providecommand \@sanitize@url [0]{\catcode `\\12\catcode `\$12\catcode
  `\&12\catcode `\#12\catcode `\^12\catcode `\_12\catcode `\%12\relax}%
\providecommand \@@startlink[1]{}%
\providecommand \@@endlink[0]{}%
\providecommand \url  [0]{\begingroup\@sanitize@url \@url }%
\providecommand \@url [1]{\endgroup\@href {#1}{\urlprefix }}%
\providecommand \urlprefix  [0]{URL }%
\providecommand \Eprint [0]{\href }%
\providecommand \doibase [0]{http://dx.doi.org/}%
\providecommand \selectlanguage [0]{\@gobble}%
\providecommand \bibinfo  [0]{\@secondoftwo}%
\providecommand \bibfield  [0]{\@secondoftwo}%
\providecommand \translation [1]{[#1]}%
\providecommand \BibitemOpen [0]{}%
\providecommand \bibitemStop [0]{}%
\providecommand \bibitemNoStop [0]{.\EOS\space}%
\providecommand \EOS [0]{\spacefactor3000\relax}%
\providecommand \BibitemShut  [1]{\csname bibitem#1\endcsname}%
\let\auto@bib@innerbib\@empty
%</preamble>
\bibitem [{\citenamefont {Yan}\ \emph {et~al.}(2013)\citenamefont {Yan},
  \citenamefont {Stein}, \citenamefont {Schaefer}, \citenamefont {Wang},\ and\
  \citenamefont {Chou}}]{yan2013}%
  \BibitemOpen
  \bibfield  {author} {\bibinfo {author} {\bibfnamefont {J.~A.}\ \bibnamefont
  {Yan}}, \bibinfo {author} {\bibfnamefont {R.}~\bibnamefont {Stein}}, \bibinfo
  {author} {\bibfnamefont {D.~M.}\ \bibnamefont {Schaefer}}, \bibinfo {author}
  {\bibfnamefont {X.~Q.}\ \bibnamefont {Wang}}, \ and\ \bibinfo {author}
  {\bibfnamefont {M.~Y.}\ \bibnamefont {Chou}},\ }\href {\doibase
  10.1103/PhysRevB.88.121403} {\bibfield  {journal} {\bibinfo  {journal} {Phys.
  Rev. B}\ }\textbf {\bibinfo {volume} {88}},\ \bibinfo {pages} {121403}
  (\bibinfo {year} {2013})}\BibitemShut {NoStop}%
\bibitem [{\citenamefont {Yan}\ \emph {et~al.}(2009)\citenamefont {Yan},
  \citenamefont {Ruan},\ and\ \citenamefont {Chou}}]{yan2009}%
  \BibitemOpen
  \bibfield  {author} {\bibinfo {author} {\bibfnamefont {J.~A.}\ \bibnamefont
  {Yan}}, \bibinfo {author} {\bibfnamefont {W.~Y.}\ \bibnamefont {Ruan}}, \
  and\ \bibinfo {author} {\bibfnamefont {M.~Y.}\ \bibnamefont {Chou}},\ }\href
  {\doibase 10.1103/PhysRevB.79.115443} {\bibfield  {journal} {\bibinfo
  {journal} {Phys. Rev. B}\ }\textbf {\bibinfo {volume} {79}},\ \bibinfo
  {pages} {115443} (\bibinfo {year} {2009})}\BibitemShut {NoStop}%
\bibitem [{\citenamefont {Piscanec}\ \emph {et~al.}(2004)\citenamefont
  {Piscanec}, \citenamefont {Lazzeri}, \citenamefont {Mauri}, \citenamefont
  {Ferrari},\ and\ \citenamefont {Robertson}}]{piscanec2004}%
  \BibitemOpen
  \bibfield  {author} {\bibinfo {author} {\bibfnamefont {S.}~\bibnamefont
  {Piscanec}}, \bibinfo {author} {\bibfnamefont {M.}~\bibnamefont {Lazzeri}},
  \bibinfo {author} {\bibfnamefont {F.}~\bibnamefont {Mauri}}, \bibinfo
  {author} {\bibfnamefont {A.~C.}\ \bibnamefont {Ferrari}}, \ and\ \bibinfo
  {author} {\bibfnamefont {J.}~\bibnamefont {Robertson}},\ }\href {\doibase
  10.1103/PhysRevLett.93.185503} {\bibfield  {journal} {\bibinfo  {journal}
  {Phys. Rev. Lett.}\ }\textbf {\bibinfo {volume} {93}},\ \bibinfo {pages}
  {185503} (\bibinfo {year} {2004})}\BibitemShut {NoStop}%
\bibitem [{\citenamefont {Lazzeri}\ \emph {et~al.}(2006)\citenamefont
  {Lazzeri}, \citenamefont {Piscanec}, \citenamefont {Mauri}, \citenamefont
  {Ferrari},\ and\ \citenamefont {Robertson}}]{lazzeri2006}%
  \BibitemOpen
  \bibfield  {author} {\bibinfo {author} {\bibfnamefont {M.}~\bibnamefont
  {Lazzeri}}, \bibinfo {author} {\bibfnamefont {S.}~\bibnamefont {Piscanec}},
  \bibinfo {author} {\bibfnamefont {F.}~\bibnamefont {Mauri}}, \bibinfo
  {author} {\bibfnamefont {A.~C.}\ \bibnamefont {Ferrari}}, \ and\ \bibinfo
  {author} {\bibfnamefont {J.}~\bibnamefont {Robertson}},\ }\href {\doibase
  10.1103/PhysRevB.73.155426} {\bibfield  {journal} {\bibinfo  {journal} {Phys.
  Rev. B}\ }\textbf {\bibinfo {volume} {73}},\ \bibinfo {pages} {155426}
  (\bibinfo {year} {2006})}\BibitemShut {NoStop}%
\bibitem [{\citenamefont {Cifuentes-Quintal}\ \emph {et~al.}(2016)\citenamefont
  {Cifuentes-Quintal}, \citenamefont {de~la {Pe\~na-Seaman}}, \citenamefont
  {Heid}, \citenamefont {de~Coss},\ and\ \citenamefont
  {Bohnen}}]{cifuentes2016}%
  \BibitemOpen
  \bibfield  {author} {\bibinfo {author} {\bibfnamefont {M.~E.}\ \bibnamefont
  {Cifuentes-Quintal}}, \bibinfo {author} {\bibfnamefont {O.}~\bibnamefont
  {de~la {Pe\~na-Seaman}}}, \bibinfo {author} {\bibfnamefont {R.}~\bibnamefont
  {Heid}}, \bibinfo {author} {\bibfnamefont {R.}~\bibnamefont {de~Coss}}, \
  and\ \bibinfo {author} {\bibfnamefont {K.-P.}\ \bibnamefont {Bohnen}},\
  }\href {\doibase 10.1103/PhysRevB.94.085401} {\bibfield  {journal} {\bibinfo
  {journal} {Phys. Rev. B}\ }\textbf {\bibinfo {volume} {94}},\ \bibinfo
  {pages} {085401} (\bibinfo {year} {2016})}\BibitemShut {NoStop}%
\bibitem [{\citenamefont {Maultzsch}\ \emph {et~al.}(2004)\citenamefont
  {Maultzsch}, \citenamefont {Reich}, \citenamefont {Thomsen}, \citenamefont
  {Requardt},\ and\ \citenamefont {Ordej\'on}}]{maultzsch2004}%
  \BibitemOpen
  \bibfield  {author} {\bibinfo {author} {\bibfnamefont {J.}~\bibnamefont
  {Maultzsch}}, \bibinfo {author} {\bibfnamefont {S.}~\bibnamefont {Reich}},
  \bibinfo {author} {\bibfnamefont {C.}~\bibnamefont {Thomsen}}, \bibinfo
  {author} {\bibfnamefont {H.}~\bibnamefont {Requardt}}, \ and\ \bibinfo
  {author} {\bibfnamefont {P.}~\bibnamefont {Ordej\'on}},\ }\href {\doibase
  10.1103/PhysRevLett.92.075501} {\bibfield  {journal} {\bibinfo  {journal}
  {Phys. Rev. Lett.}\ }\textbf {\bibinfo {volume} {92}},\ \bibinfo {pages}
  {075501} (\bibinfo {year} {2004})}\BibitemShut {NoStop}%
\bibitem [{Note1()}]{Note1}%
  \BibitemOpen
  \bibinfo {note} {At least in absence of spin-orbit coupling, as was
  considered in the article of Yan \protect \textit {et al.}\cite
  {yan2013}}\BibitemShut {NoStop}%
\bibitem [{\citenamefont {Giannozzi}\ \emph {et~al.}(2009)\citenamefont
  {Giannozzi}, \citenamefont {Baroni}, \citenamefont {Bonini}, \citenamefont
  {Calandra}, \citenamefont {Car}, \citenamefont {Cavazzoni}, \citenamefont
  {Ceresoli}, \citenamefont {Chiarotti}, \citenamefont {Cococcioni},
  \citenamefont {Dabo}, \citenamefont {{Dal Corso}}, \citenamefont
  {de~Gironcoli}, \citenamefont {Fabris}, \citenamefont {Fratesi},
  \citenamefont {Gebauer}, \citenamefont {Gerstmann}, \citenamefont
  {Gougoussis}, \citenamefont {Kokalj}, \citenamefont {Lazzeri}, \citenamefont
  {Martin-Samos}, \citenamefont {Marzari}, \citenamefont {Mauri}, \citenamefont
  {Mazzarello}, \citenamefont {Paolini}, \citenamefont {Pasquarello},
  \citenamefont {Paulatto}, \citenamefont {Sbraccia}, \citenamefont {Scandolo},
  \citenamefont {Sclauzero}, \citenamefont {Seitsonen}, \citenamefont
  {Smogunov}, \citenamefont {Umari},\ and\ \citenamefont
  {Wentzcovitch}}]{QE-2009}%
  \BibitemOpen
  \bibfield  {author} {\bibinfo {author} {\bibfnamefont {P.}~\bibnamefont
  {Giannozzi}}, \bibinfo {author} {\bibfnamefont {S.}~\bibnamefont {Baroni}},
  \bibinfo {author} {\bibfnamefont {N.}~\bibnamefont {Bonini}}, \bibinfo
  {author} {\bibfnamefont {M.}~\bibnamefont {Calandra}}, \bibinfo {author}
  {\bibfnamefont {R.}~\bibnamefont {Car}}, \bibinfo {author} {\bibfnamefont
  {C.}~\bibnamefont {Cavazzoni}}, \bibinfo {author} {\bibfnamefont
  {D.}~\bibnamefont {Ceresoli}}, \bibinfo {author} {\bibfnamefont {G.~L.}\
  \bibnamefont {Chiarotti}}, \bibinfo {author} {\bibfnamefont {M.}~\bibnamefont
  {Cococcioni}}, \bibinfo {author} {\bibfnamefont {I.}~\bibnamefont {Dabo}},
  \bibinfo {author} {\bibfnamefont {A.}~\bibnamefont {{Dal Corso}}}, \bibinfo
  {author} {\bibfnamefont {S.}~\bibnamefont {de~Gironcoli}}, \bibinfo {author}
  {\bibfnamefont {S.}~\bibnamefont {Fabris}}, \bibinfo {author} {\bibfnamefont
  {G.}~\bibnamefont {Fratesi}}, \bibinfo {author} {\bibfnamefont
  {R.}~\bibnamefont {Gebauer}}, \bibinfo {author} {\bibfnamefont
  {U.}~\bibnamefont {Gerstmann}}, \bibinfo {author} {\bibfnamefont
  {C.}~\bibnamefont {Gougoussis}}, \bibinfo {author} {\bibfnamefont
  {A.}~\bibnamefont {Kokalj}}, \bibinfo {author} {\bibfnamefont
  {M.}~\bibnamefont {Lazzeri}}, \bibinfo {author} {\bibfnamefont
  {L.}~\bibnamefont {Martin-Samos}}, \bibinfo {author} {\bibfnamefont
  {N.}~\bibnamefont {Marzari}}, \bibinfo {author} {\bibfnamefont
  {F.}~\bibnamefont {Mauri}}, \bibinfo {author} {\bibfnamefont
  {R.}~\bibnamefont {Mazzarello}}, \bibinfo {author} {\bibfnamefont
  {S.}~\bibnamefont {Paolini}}, \bibinfo {author} {\bibfnamefont
  {A.}~\bibnamefont {Pasquarello}}, \bibinfo {author} {\bibfnamefont
  {L.}~\bibnamefont {Paulatto}}, \bibinfo {author} {\bibfnamefont
  {C.}~\bibnamefont {Sbraccia}}, \bibinfo {author} {\bibfnamefont
  {S.}~\bibnamefont {Scandolo}}, \bibinfo {author} {\bibfnamefont
  {G.}~\bibnamefont {Sclauzero}}, \bibinfo {author} {\bibfnamefont {A.~P.}\
  \bibnamefont {Seitsonen}}, \bibinfo {author} {\bibfnamefont {A.}~\bibnamefont
  {Smogunov}}, \bibinfo {author} {\bibfnamefont {P.}~\bibnamefont {Umari}}, \
  and\ \bibinfo {author} {\bibfnamefont {R.~M.}\ \bibnamefont {Wentzcovitch}},\
  }\href {http://stacks.iop.org/0953-8984/21/i=39/a=395502} {\bibfield
  {journal} {\bibinfo  {journal} {J. Phys.: Condens. Matter}\ }\textbf
  {\bibinfo {volume} {21}},\ \bibinfo {pages} {395502} (\bibinfo {year}
  {2009})}\BibitemShut {NoStop}%
\bibitem [{\citenamefont {Perdew}\ and\ \citenamefont {Zunger}(1981)}]{PZ}%
  \BibitemOpen
  \bibfield  {author} {\bibinfo {author} {\bibfnamefont {J.-P.}\ \bibnamefont
  {Perdew}}\ and\ \bibinfo {author} {\bibfnamefont {A.}~\bibnamefont
  {Zunger}},\ }\href {\doibase 10.1103/PhysRevB.23.5048} {\bibfield  {journal}
  {\bibinfo  {journal} {Phys. Rev. B}\ }\textbf {\bibinfo {volume} {23}},\
  \bibinfo {pages} {5048} (\bibinfo {year} {1981})}\BibitemShut {NoStop}%
\bibitem [{\citenamefont {Corso}(2014)}]{PSLIB}%
  \BibitemOpen
  \bibfield  {author} {\bibinfo {author} {\bibfnamefont {A.~D.}\ \bibnamefont
  {Corso}},\ }\href {\doibase
  http://dx.doi.org/10.1016/j.commatsci.2014.07.043} {\bibfield  {journal}
  {\bibinfo  {journal} {Comput. Mater. Sci}\ }\textbf {\bibinfo {volume}
  {95}},\ \bibinfo {pages} {337 } (\bibinfo {year} {2014})}\BibitemShut
  {NoStop}%
\bibitem [{\citenamefont {Methfessel}\ and\ \citenamefont {Paxton}(1989)}]{MP}%
  \BibitemOpen
  \bibfield  {author} {\bibinfo {author} {\bibfnamefont {M.}~\bibnamefont
  {Methfessel}}\ and\ \bibinfo {author} {\bibfnamefont {A.~T.}\ \bibnamefont
  {Paxton}},\ }\href {\doibase 10.1103/PhysRevB.40.3616} {\bibfield  {journal}
  {\bibinfo  {journal} {Phys. Rev. B}\ }\textbf {\bibinfo {volume} {40}},\
  \bibinfo {pages} {3616} (\bibinfo {year} {1989})}\BibitemShut {NoStop}%
\bibitem [{ELK()}]{ELK}%
  \BibitemOpen
  \href@noop {} {\bibinfo  {journal} {http://elk.sourceforge.net}\
  }\BibitemShut {NoStop}%
\bibitem [{\citenamefont {{De la Pe\~na}-Seaman}\ \emph
  {et~al.}(2007)\citenamefont {{De la Pe\~na}-Seaman}, \citenamefont {de~Coss},
  \citenamefont {Heid},\ and\ \citenamefont {Bohnen}}]{omar2007}%
  \BibitemOpen
\bibfield  {journal} {  }\bibfield  {author} {\bibinfo {author} {\bibfnamefont
  {O.}~\bibnamefont {{De la Pe\~na}-Seaman}}, \bibinfo {author} {\bibfnamefont
  {R.}~\bibnamefont {de~Coss}}, \bibinfo {author} {\bibfnamefont
  {R.}~\bibnamefont {Heid}}, \ and\ \bibinfo {author} {\bibfnamefont {K.~P.}\
  \bibnamefont {Bohnen}},\ }\href {\doibase 10.1103/PhysRevB.76.174205}
  {\bibfield  {journal} {\bibinfo  {journal} {Phys. Rev. B}\ }\textbf {\bibinfo
  {volume} {76}},\ \bibinfo {pages} {174205} (\bibinfo {year}
  {2007})}\BibitemShut {NoStop}%
\bibitem [{\citenamefont {Mart\'inez-Guerra}\ \emph {et~al.}(2014)\citenamefont
  {Mart\'inez-Guerra}, \citenamefont {Ort\'iz-Chi}, \citenamefont {Curtarolo},\
  and\ \citenamefont {de~Coss}}]{edgar}%
  \BibitemOpen
  \bibfield  {author} {\bibinfo {author} {\bibfnamefont {E.}~\bibnamefont
  {Mart\'inez-Guerra}}, \bibinfo {author} {\bibfnamefont {F.}~\bibnamefont
  {Ort\'iz-Chi}}, \bibinfo {author} {\bibfnamefont {S.}~\bibnamefont
  {Curtarolo}}, \ and\ \bibinfo {author} {\bibfnamefont {R.}~\bibnamefont
  {de~Coss}},\ }\href {http://stacks.iop.org/0953-8984/26/i=11/a=115701}
  {\bibfield  {journal} {\bibinfo  {journal} {J. Phys.: Condens. Matter}\
  }\textbf {\bibinfo {volume} {26}},\ \bibinfo {pages} {115701} (\bibinfo
  {year} {2014})}\BibitemShut {NoStop}%
\bibitem [{\citenamefont {Lazzeri}\ \emph {et~al.}(2008)\citenamefont
  {Lazzeri}, \citenamefont {Attaccalite}, \citenamefont {Wirtz},\ and\
  \citenamefont {Mauri}}]{lazzeri2008}%
  \BibitemOpen
  \bibfield  {author} {\bibinfo {author} {\bibfnamefont {M.}~\bibnamefont
  {Lazzeri}}, \bibinfo {author} {\bibfnamefont {C.}~\bibnamefont
  {Attaccalite}}, \bibinfo {author} {\bibfnamefont {L.}~\bibnamefont {Wirtz}},
  \ and\ \bibinfo {author} {\bibfnamefont {F.}~\bibnamefont {Mauri}},\ }\href
  {\doibase 10.1103/PhysRevB.78.081406} {\bibfield  {journal} {\bibinfo
  {journal} {Phys. Rev. B}\ }\textbf {\bibinfo {volume} {78}},\ \bibinfo
  {pages} {081406} (\bibinfo {year} {2008})}\BibitemShut {NoStop}%
\end{thebibliography}
%

\end{document}